\RequirePackage[2020-02-02]{latexrelease} 
\documentclass[%
 reprint,
 amsmath,amssymb,
 aps,
]{revtex4-1}

\usepackage{graphicx}
\usepackage{dcolumn}
\usepackage{bm}
\usepackage{lipsum}
\usepackage{amsfonts}
\usepackage{amssymb}
\usepackage{textcomp}
\begin{document}



\title{Nonlinearity-Inhomogeneity Competition in Discrete-Time Quantum Walks}

\author{N. Amaral$^1$, A. R. C. Buarque$^2$ and W. S. Dias$^1$}
\address{$^1$Instituto de F\' isica, Universidade Federal de Alagoas, 57072-900 Macei\' o, Alagoas, Brazil.}
\address{$^2$Quantum Industrial Innovation, Centro de Competência Embrapii Cimatec, SENAI CIMATEC,
Av. Orlando Gomes, 1845 Salvador-BA, Brazil}

\begin{abstract}
We investigate the interplay between nonlinearity and inhomogeneities in discrete-time quantum walks on one-dimensional lattices. Nonlinear effects are introduced through a Kerr-like, intensity-dependent local phase, while spatial and temporal inhomogeneities are implemented via random variations of the quantum gate operations. By analyzing typical quantities, such as the return probability and the participation function, we identify distinct quantum walking regimes as the nonlinear parameter $\chi$ and the quantum gate parameter $\theta$ are varied. Spatial inhomogeneities weaken nonlinear self-trapping and constrict the region of robust localization. In this process, partially localized regimes emerge, characterized by the coexistence of a confined core and dispersive wave-packet components. In contrast, temporal inhomogeneities act as time-dependent perturbations that continuously disrupt the phase coherence required for self-trapped excitation, thereby enhancing dispersive emission and promoting delocalization. By using $\chi$ versus $\theta$ diagrams, we display a comprehensive characterization of how inhomogeneities modify the stability and extent of prevailing dynamical regimes, elucidating the competition between nonlinearity and inhomogeneities in discrete-time quantum walks. \end{abstract}


\maketitle

\section{introduction}
\label{introduction}

Quantum walks on lattices have become a significant framework for quantum information processing and simulation, providing an intuitive and versatile platform to explore coherent transport phenomena, interference effects, and algorithmic speedups when compared to their classical counterparts~\cite{PhysRevLett.102.180501,10.1145/380752.380758, PhysRevA.70.022314, Venegas-Andraca2012, Xu2021}. In particular, discrete-time quantum walks (DTQWs) have been extensively investigated due to their experimental accessibility in a variety of platforms, including photonic systems~\cite{PhysRevLett.106.180403, PhysRevLett.122.020501}, ultracold atoms~\cite{PhysRevLett.124.050502,doi:10.1126/science.1260364}, trapped ions~\cite{HuertaAlderete2020,Bruzewicz2019}, and superconducting circuits~\cite{5lm1-2kpk}. Beyond their experimental relevance, DTQWs provide a fully controllable setting to explore coherent transport and interference phenomena.

While the standard formulation of DTQWs relies on linear and unitary evolution, a growing body of research has shown that effective nonlinearities modify the coherent dynamics, giving rise to a rich phenomenology that departs from purely ballistic spreading. In this context, nonlinear DTQWs have been shown to exhibit slow diffusion~\cite{Shikano2014}, self-trapping~\cite{PhysRevA.101.023802, PhysRevA.106.062407}, breathing dynamics~\cite{PhysRevA.103.042213}, and soliton-like formations~\cite{PhysRevA.75.062333, PhysRevA.106.042202,PhysRevA.101.062335}.

Such phenomena depend sensitively on both the nonlinear strength and the specific choice of quantum gates~\cite{PhysRevA.101.023802, PhysRevA.106.062407, PhysRevA.75.062333, PhysRevA.106.042202, PhysRevA.101.062335, PhysRevA.103.042213}. They may even display counterintuitive transitions in which increasing the nonlinearity induces delocalization of the wave packet rather than enhanced localization~\cite{PhysRevA.101.023802, PhysRevA.106.062407}. In this sense, previous studies have also shown that qualitatively different dynamical regimes emerge depending on how the nonlinearity is implemented, whether through state-dependent quantum-coin operations~\cite{Shikano2014}, Kerr-like intensity-dependent phase shifts~\cite{PhysRevA.75.062333, PhysRevA.106.062407, PhysRevA.101.023802, PhysRevA.106.042202, PhysRevA.103.042213}, or nonlinear displacement mechanisms~\cite{PhysRevA.101.062335}. Taken together, these features raise the question of how robust such nonlinear regimes are under spatial or temporal variations of the evolution operators.

Achieving ideal homogeneous conditions over long time scales is uncommon in practical implementations. In linear DTQWs, spatial or temporal inhomogeneities in the underlying evolution protocol may arise from fluctuations in the coin operator, site-dependent phase shifts, uneven tunneling amplitudes, or variations in the conditional displacement. These inhomogeneities significantly modify the coherent dynamics, leading to either diffusive behavior~\cite{PhysRevLett.106.180403, PhysRevA.74.022310, PhysRevA.68.062315} or localization phenomena~\cite{PhysRevLett.106.180403, PhysRevA.89.022309, PhysRevA.89.042307}, depending on their nature. Other transport regimes, including sub- or superdiffusive behavior, may also emerge depending on additional features, such as correlated inhomogeneities~\cite{PhysRevE.99.022117, PhysRevE.109.064151}, aperiodicity~\cite{PhysRevE.96.012111, PhysRevE.100.032106}, or fractal patterns~\cite{doi:10.1142/S021974991350069X,PhysRevA.109.022217}. Extending these analyses, we observe only a few studies that incorporate both nonlinearity and inhomogeneities in DTQWs~\cite{PhysRevLett.122.040501,PhysRevA.105.L020202}. Using a unitary map approach, it was shown that nonlinearity breaks Anderson localization and that the subdiffusive scaling reported for Gross-Pitaevskii lattices extends over four additional decades in time~\cite{PhysRevLett.122.040501}. Within the same framework, corrections to the nonlinear evolution were shown to induce an intermediate logarithmic expansion regime between the mean-field subdiffusion and the final saturation~\cite{PhysRevA.105.L020202}.

Despite these advances, it remains unclear whether spatial or temporal inhomogeneities preserve, destabilize, or qualitatively transform the nonlinear dynamical behaviors established for homogeneous DTQWs, including chaotic-like, travelling self-trapped (soliton-like), and stationary self-trapped regimes~\cite{PhysRevA.75.062333,PhysRevA.101.023802}. In this work, we investigate the interplay between nonlinearity and inhomogeneities in discrete-time quantum walks on one-dimensional lattices. Nonlinear effects are introduced through a Kerr-like, intensity-dependent local phase associated with the walker along its propagation~\cite{PhysRevA.75.062333,PhysRevA.101.023802}, while spatial and temporal inhomogeneities are implemented via random variations of the quantum gate operations around a reference parameter $\theta_0$. Taking the homogeneous scenario as a benchmark, we analyze how such inhomogeneities modify the stability and persistence of these established nonlinear regimes. We analyze the return probability and the participation function, constructing $(\chi,\theta_0)$ diagrams to map the dynamical behavior across parameter space. Our results reveal the emergence of partially localized regimes and inhomogeneity-induced spreading mechanisms that are absent in homogeneous lattices. These findings elucidate the competition between nonlinearity and inhomogeneities in discrete-time quantum walks.

\section{model}
\label{model}

We consider discrete-time quantum walks on one-dimensional lattices composed of interconnected sites. The walker is modeled as a two-level system whose state $|\psi\rangle$ belongs to the composite Hilbert space $\mathcal{H} = \mathcal{H}_P \otimes \mathcal{H}_C$.
$\mathcal{H}_C$ is a two-dimensional complex space associated with the internal degree of freedom (the coin space), spanned by the basis $\{|R\rangle, |L\rangle\}$. The position Hilbert space $\mathcal{H}_P$ is spanned by the set of orthonormal states $\{|n\rangle\}$,
with $n \in \mathbb{Z}$ labeling the lattice sites. Thus, the state of the quantum walker at the $t$th time step can be written as
\begin{equation}
|\psi(t)\rangle = \sum_n \left[a_{n,t}|R\rangle + b_{n,t}|L\rangle\right]\otimes |n\rangle,
\end{equation}
where the complex amplitudes $a_{n,t}$ and $b_{n,t}$ satisfy the normalization condition
$\sum_n \left(|a_{n,t}|^2 + |b_{n,t}|^2\right) = 1$.

Each step of the linear quantum walk consists of two basic operations. First, local quantum gates $\hat{C}$ act on the internal degree of freedom of the walker at each lattice site, shuffling its internal state. Then, the conditional shift operator $\hat{S}$ redistributes the probability amplitudes along the lattice according to the internal state. Starting from an initial state $|\psi(t=0)\rangle$, the linear evolution proceeds recursively as $|\psi(t)\rangle = \hat U_L |\psi(t-1)\rangle$, where the single-step linear evolution operator is given by $\hat U_L = \hat S \, (\hat C \otimes \hat I_P)$, with $\hat I_P$ denoting the identity operator on the position Hilbert space.

The quantum gates $\hat{C}$ are unitary operators parameterized as
\begin{equation}\label{quantum_coin_operator}
\hat{C}(\theta) = \cos(\theta) \, \hat{Z} + \sin(\theta) \, \hat{X},
\end{equation}
where $\theta$ controls the spreading of the walker~\cite{PhysRevA.67.052307}, and $\hat{Z}$ and $\hat{X}$ denote the Pauli matrices. The spatial redistribution of probability amplitudes is implemented by the conditional shift operator
\begin{small}
\begin{equation}
\hat{S} = \sum_n \left( |n+1\rangle\langle n| \otimes |R\rangle\langle R| 
+ |n-1\rangle\langle n| \otimes |L\rangle\langle L| \right),
\end{equation}
\end{small}
which couples the internal state of the walker to its spatial displacement along the lattice.

Nonlinear effects are incorporated by supplementing the linear protocol with a local phase operator
$\hat K^t$, which accounts for an intensity-dependent phase acquired by each internal component at every lattice site~\cite{PhysRevA.75.062333,PhysRevA.101.023802}. Thus, the full nonlinear evolution proceeds recursively as
\begin{equation}
|\psi(t)\rangle = \hat U^t |\psi(t-1)\rangle,
\end{equation}
with $\hat U^t = \hat U_L \hat K^{t-1}$. The nonlinear phase operator is defined as
\begin{small}
\begin{align}
\hat K^{t}
&= \sum_{s=R,L}\sum_{n} e^{i G^t(n,s)} |s\rangle\langle s|\otimes |n\rangle\langle n| \\
&= \sum_{n} \left(e^{i G^t(n,R)}|R\rangle\langle R| + e^{i G^t(n,L)}|L\rangle\langle L|\right) \otimes |n\rangle\langle n|,
\end{align}
\end{small}
where $G^t(n,s)$ is a real-valued function of the local probability amplitudes,
which may depend on the lattice site $n$, the internal state $s$, and the time step $t$~\cite{PhysRevA.75.062333,PhysRevA.101.023802}. By considering a Kerr-like nonlinearity, we set $G^t(n,s)=2\pi\chi\,|\psi^t_{n,s}|^2$,
where $|\psi^t_{n,s}|^2=|\langle n,s|\psi(t)\rangle|^2$ denotes the local probability associated with the internal state $s$ at site $n$ and time step $t$. The parameter $\chi$ quantifies the strength of the nonlinear response, and the linear discrete-time quantum walk is recovered in the limit $\chi=0$.

Building on this nonlinear framework, we introduce controlled spatial or temporal inhomogeneities in the quantum gate operations. Starting from a homogeneous reference value $\theta_0$, the coin parameter is allowed to undergo random variations, such that
 \begin{equation} \theta \rightarrow \theta_0 + \delta, \end{equation}
  where $\delta$ denotes independent, zero-mean random variables uniformly distributed in the interval $[-W/2, W/2]$, with variance $\mathrm{Var}[\delta] = W^2/12$. The homogeneous scenario is recovered for $\delta=0$, yielding $\theta=\theta_0$. Spatial inhomogeneities correspond to site-dependent coin parameters $\theta(n)=\theta_0+\delta_n$~\cite{PhysRevLett.106.180403,PhysRevE.100.032106}, whereas temporal inhomogeneities are modeled by step-dependent variations $\theta(t)=\theta_0+\delta_t$~\cite{PhysRevLett.106.180403,PhysRevE.109.064151}. The inhomogeneity strength mainly influences the transient dynamics preceding the asymptotic regime. The choice $W=10$ throughout this work allows for a straightforward numerical characterization of these effects within feasible simulation times, while smaller values do not qualitatively alter the long-time behavior.

We consider the initial state of the quantum walker to be a
symmetric one of the form
\begin{equation}
|\psi(t=0)\rangle = \dfrac{1}{\sqrt{2}}(|R\rangle + i|L \rangle)\otimes |n_0\rangle,
\label{def:inistate}
\end{equation}
with the initial position $n_0$ of the quantum walker defined at the central site of the lattice. Open boundary conditions are employed throughout the analysis, using sufficiently large lattice sizes to prevent the wave function from reaching the edges during the specified time course. Considering the stochastic nature and the uniqueness of each sample owing to its specific inhomogeneity, averaging over multiple samples provides a more representative and robust perspective on the system. Specifically, we establish an ensemble of 50 independent quantum walks to evaluate its average behavior.

\section{Results and discussion}
\label{Results_and_discussion}  

\begin{figure}[t]
    \centering
    \includegraphics[height=8.9cm]{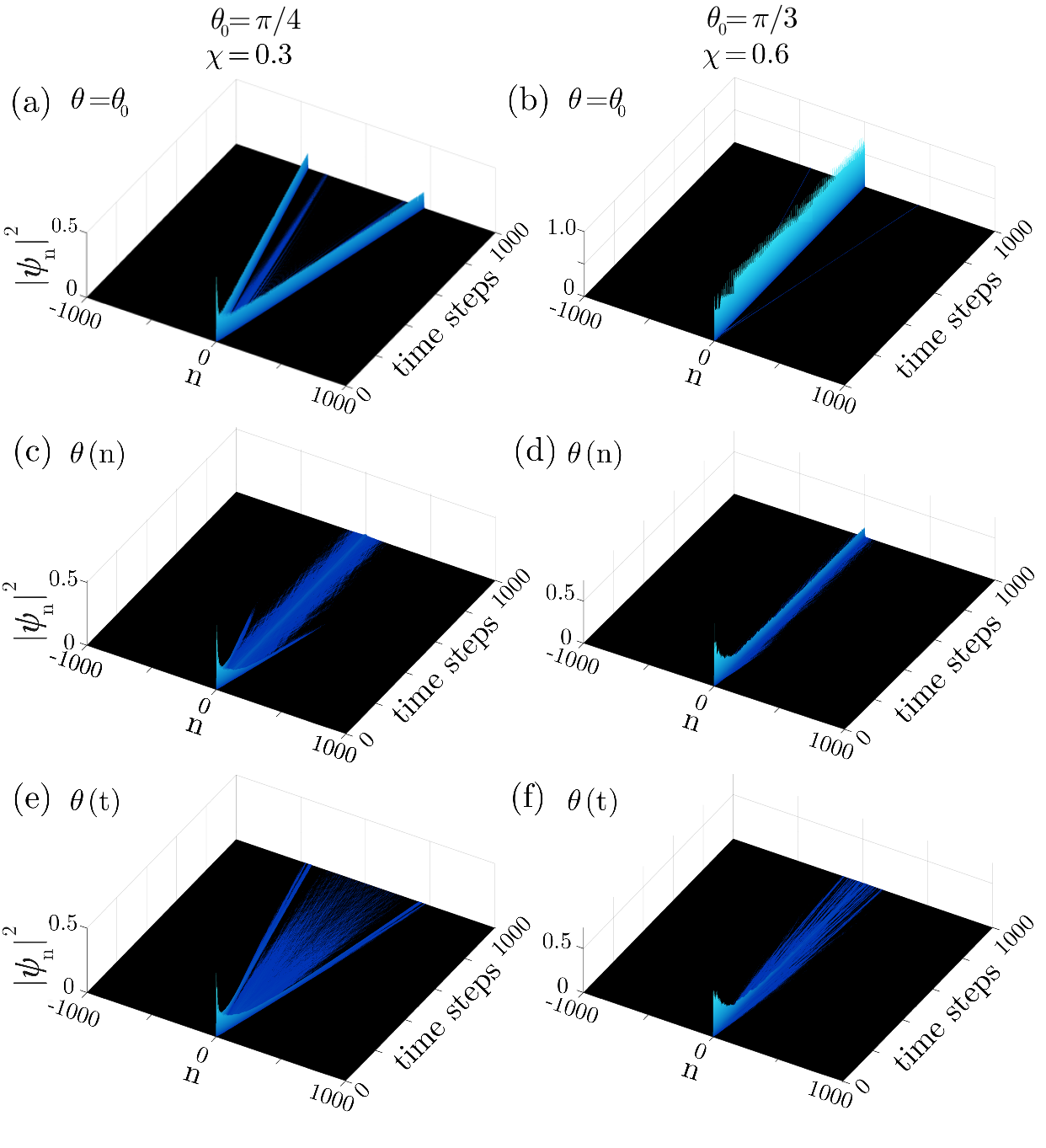}
    \caption{(Color on-line) Time evolution of the position-space probability density of a quantum walker on one-dimensional nonlinear lattices. The left (right) column corresponds to $\theta_0=\pi/4$, $\chi=0.3$ ($\theta_0=\pi/3$, $\chi=0.6$). The homogeneous scenarios [(a),(b)] exhibit solitonlike and stationary self-trapped regimes, in full agreement with the results reported in Refs.~\cite{PhysRevA.75.062333, PhysRevA.101.023802}. Spatial [(c),(d)] and temporal [(e),(f)] inhomogeneities suppress the nonlinear coherent dynamics responsible for the formation of traveling and stationary self-trapped states.} 
    \label{fig:1}
\end{figure}

We start by analyzing the time evolution of the position-space probability density \( |\psi_n(t)|^2 \) for representative values of the nonlinear parameter \(\chi\). The walker is initially prepared in a balanced superposition of left- and right-handed circular states (see Eq.~\ref{def:inistate}), with its initial position localized at the lattice center. Fig.~\ref{fig:1} shows the resulting dynamics for chains ruled by quantum gates characterized by \(\theta_0=\pi/4\) with nonlinear strength \(\chi=0.3\) (left column) and by \(\theta_0=\pi/3\) with \(\chi=0.6\) (right column). Panels [(a),(b)] correspond to homogeneously distributed quantum gates, panels [(c),(d)] to spatially inhomogeneous gates \(\theta(n)\), and panels [(e),(f)] to temporally inhomogeneous gates \(\theta(t)\). In the homogeneous scenario (\(\theta=\theta_0\)), the dynamics display the well-known travelling and stationary self-trapped regimes, in full agreement with Refs.~\cite{PhysRevA.75.062333, PhysRevA.101.023802}. When spatial disorder of strength is introduced [(c),(d)], the plots suggest the emergence of localized features, with a fraction of the wave packet remaining trapped around the initial site. In contrast, temporal inhomogeneities weaken self-trapped and localized features, promoting more extended and spatially widespread regimes [(e),(f)]. Overall, such results suggest that inhomogeneities suppress the nonlinear coherent dynamics responsible for the formation of both traveling and stationary self-trapped states.

In order to further characterize the dynamical regimes observed in Fig.~\ref{fig:1}, we explore quantities that allow us to probe the long-time behavior of the quantum walk. In particular, we compute the participation function \(\mathrm{PR}(t)\),
\begin{equation}
\mathrm{PR}(t) = \left[ \sum_n |\psi_n(t)|^4 \right]^{-1},
\end{equation}
and the return probability \(\mathrm{R_0}(t)\),
\begin{equation}
\mathrm{R_0}(t) = |\psi_{n_0}(t)|^2,
\end{equation}
for the same parameter configurations. The participation function provides an estimate of the effective number of lattice sites significantly populated by the wave packet at time $t$, while \(\mathrm{R_0}(t)\) measures the probability of finding the walker at the initial position. In the long-time regime, a saturation of $\mathrm{R_0}(t)$ at a finite value signals localization, while $\mathrm{R_0}(t)\to 0$ indicates that the walker progressively escapes from the initial site.

\begin{figure}[t]
    \centering
    \includegraphics[height=7.9cm]{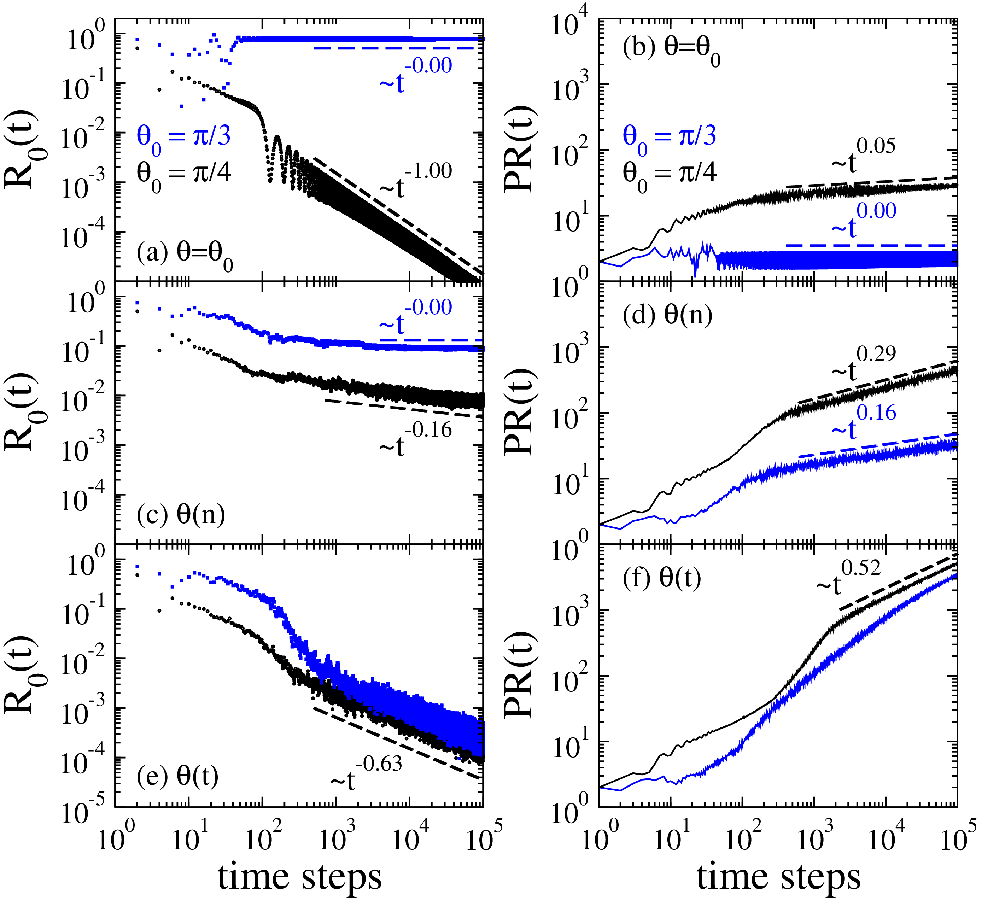}
    \caption{(Color on-line) Time evolution of the return probability $\mathrm{R_0}(t)$ and the participation function $\mathrm{PR}(t)$ for the same configurations as in Fig.~\ref{fig:1}. Spatial [(c),(d)] and temporal inhomogeneities [(e),(f)] of the quantum gates suppress the coherent nonlinear dynamics underlying travelling and stationary self-trapped regimes.} 
    \label{fig:2}
\end{figure}


In Fig.~\ref{fig:2}, we examine the time evolution of the return probability $\mathrm{R_0}(t)$ (left column) and the participation function $\mathrm{PR}(t)$ (right column) for the same parameter configurations shown in Fig.~\ref{fig:1}, now followed over a longer time evolution.  In panels (a) and (b), corresponding to homogeneous lattices, the configuration with $\theta_0=\pi/4$ and $\chi=0.3$ exhibits a decay of the return probability following $\mathrm{R_0}(t)\propto t^{-1}$. At the same time, the participation function stays small and approximately constant over time, a behavior consistent with travelling self-trapped modes~\cite{PhysRevA.75.062333, PhysRevA.101.023802}. Conversely, for $\theta_0=\pi/3$ and $\chi=0.6$, $\mathrm{R_0}(t)$ remains close to unity, while $\mathrm{PR}(t)$ saturates after an initial transient, corroborating the stationary self-trapped regime~\cite{PhysRevA.101.023802}.

When spatial inhomogeneities are introduced [panels (c) and (d)], the static disorder locally modifies the effective parameters of the internal degree of freedom $\{|R\rangle, |L\rangle\}$, inducing partial backscattering and mode conversion that progressively degrade the phase coherence required for the formation and sustained existence of a self-trapped excitation. As a result, the travelling self-trapped regime ($\theta_0=\pi/4$ with $\chi=0.3$) gives way to a slower decay of the return probability, $\mathrm{R_0}(t)\propto t^{-0.16}$, accompanied by a subdiffusive growth of the participation function, $\mathrm{PR}(t)\propto t^{0.29}$. These features are associated with the emission of dispersive wave-packet components, coexisting with a residual probability density concentrated around the initial position that progressively diminishes as time evolves~\cite{PhysRevLett.70.1787,PhysRevB.58.12547}. In contrast, for the initially stationary self-trapped regime ($\theta_0=\pi/3$ with $\chi=0.6$), a finite fraction of the probability density remains pinned near the initial site, while the remaining portion gradually escapes and spreads through the lattice, leading to $\mathrm{PR}(t)\propto t^{0.16}$. This coexistence of a long-lived localized core with simultaneously spreading components differs from the purely radiative decay commonly discussed in nonlinear disordered lattice systems, which suggests a distinct dynamical regime enabled by the interplay between strong nonlinearity and spatial inhomogeneity in nonlinear discrete-time quantum walks.

For temporal inhomogeneities [panels (e) and (f)], both initially travelling and stationary configurations converge, after an initial transient, to the same asymptotic behavior. This regime exhibits a decay of the return probability, $\mathrm{R_0}(t)\propto t^{-0.63}$, steeper than in the spatially inhomogeneous lattice. The diffusive growth of the participation function, $\mathrm{PR}(t)\propto t^{0.52}$, signals a substantially enhanced wave-packet spreading. This behavior arises from the fact that temporal inhomogeneities act as time-dependent perturbations that continuously reshape the interference conditions of the walk. Such dynamical disorder inhibits the formation of persistent resonant structures and efficiently breaks the phase coherence required to sustain nonlinear self-trapping. As a consequence, self-trapping is suppressed, and the long-time dynamics resemble an effectively diffusive-like spreading driven by temporal inhomogeneities~\cite{PhysRevLett.106.180403,PhysRevA.89.042307,PhysRevE.109.064151}.

\begin{figure}[t]
    \centering
    \includegraphics[height=8.6cm]{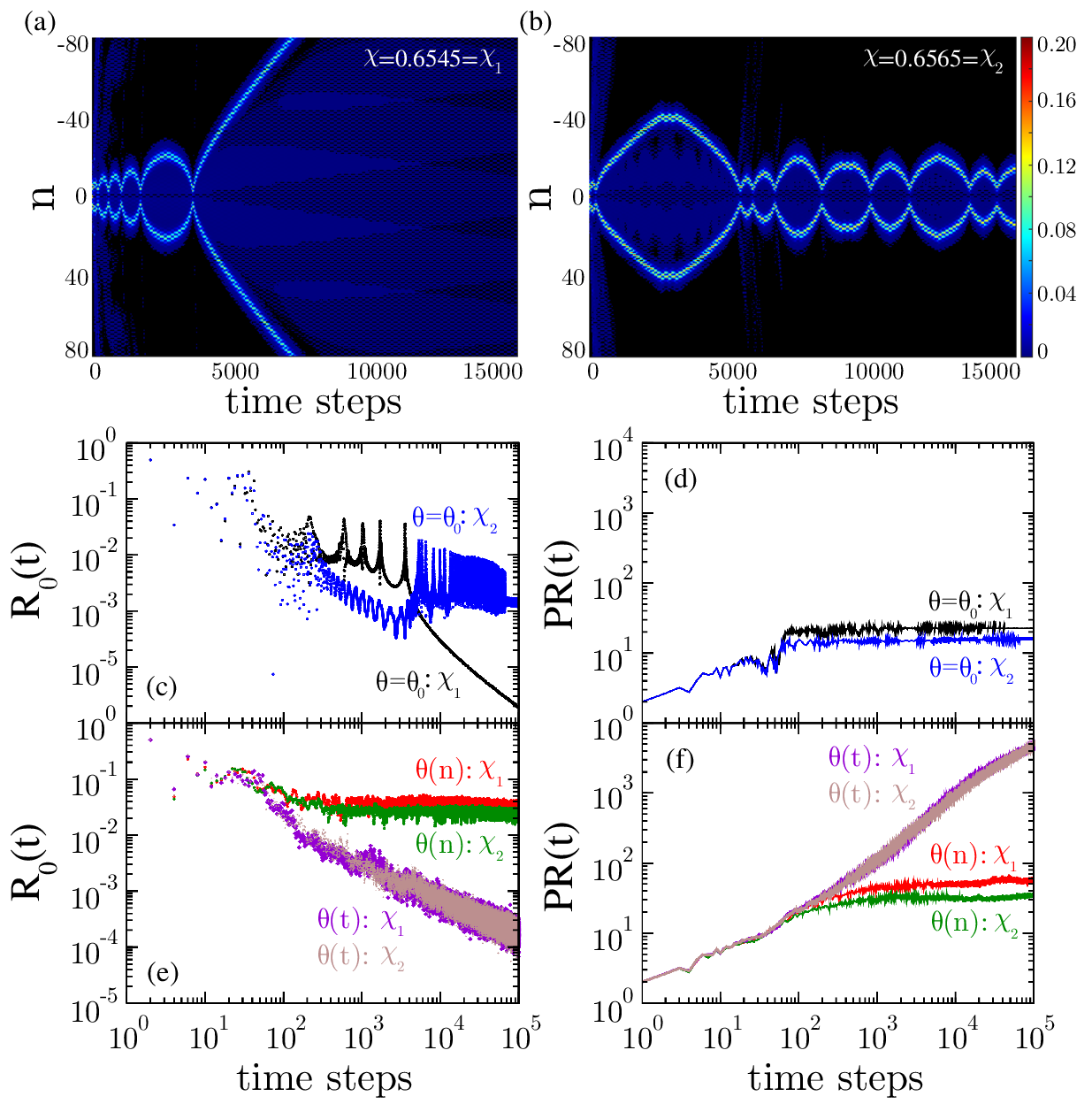}
    \caption{(Color on-line)  In the absence of inhomogeneities, the position-space probability density exhibits emerging self-trapped profiles that become extremely sensitive to minor variations of the nonlinear strength $\chi$ [(a),(b)]. The corresponding return probability $\mathrm{R_0}(t)$ (c) and participation function $\mathrm{PR}(t)$ (d) reveal wave packets either remaining confined and oscillating around the origin or eventually escaping from the initial position. Spatial and temporal inhomogeneities [(e),(f)] suppress this chaotic-like dynamics, promoting localized and extended regimes, respectively.} 
    \label{fig:3}
\end{figure}

\begin{figure*}[t]
    \centering
    \includegraphics[height=9.6cm]{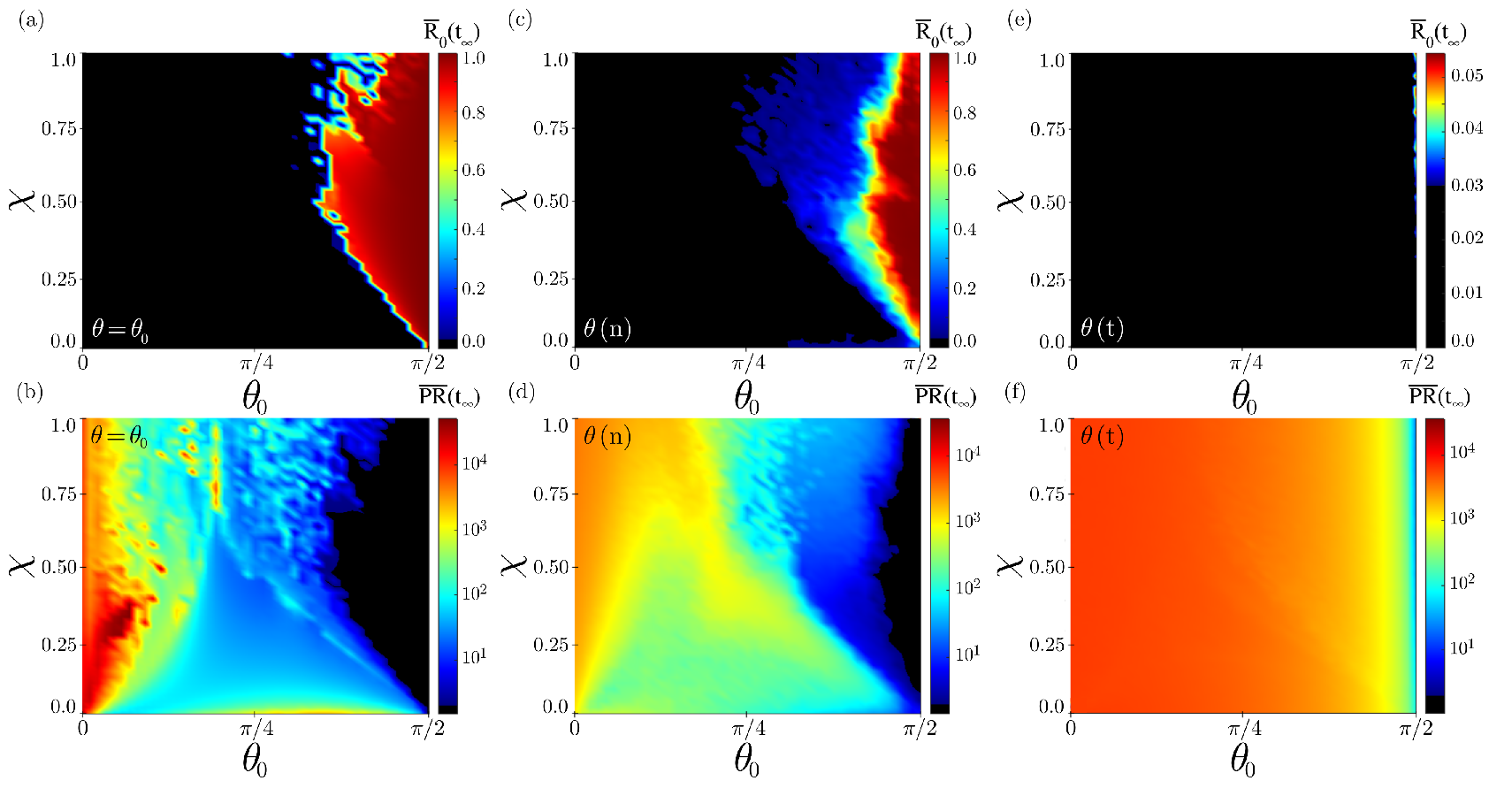}
    \caption{(Color on-line)  Long-time averaged return probability (top panels) and participation function (bottom panels) mapped in the $(\chi,\theta_0)$ plane for (a,b) homogeneous lattices and for lattices with (c,d) spatial and (e,f) temporal inhomogeneities in the quantum gates. In the homogeneous scenario, stationary self-trapping is favored as $\theta_0$ approaches $\pi/2$, in agreement with Ref.~\cite{PhysRevA.101.023802}. Spatial inhomogeneities substantially reduce the parameter region associated with strong stationary self-trapping, leading to the emergence of partially localized states. In contrast, temporal inhomogeneities entirely suppress localized regimes, promoting extended asymptotic dynamics.} 
    \label{fig:4}
\end{figure*}

Another dynamical regime reported in Refs.~\cite{PhysRevA.75.062333, PhysRevA.101.023802} concerns the emergence of travelling self-trapped formations that are extremely sensitive to small variations of the nonlinear parameter $\chi$. Figs.~\ref{fig:3}(a) and (b) illustrate representative examples of this regime, showing the time evolution of the position-space probability density for a quantum walker evolving on a homogeneous lattice of Hadamard quantum gates ($\theta=\pi/4$) with nonlinearities (a) $\chi=0.6545 = \chi_1$ and (b) $\chi=0.6565 = \chi_2$. Despite the minute difference between these values, the resulting dynamics are markedly distinct, ranging from travelling self-trapped formations that progressively move away from the initial position to regimes in which such formations remain confined and oscillate around it. This pronounced sensitivity to the nonlinear coupling corroborates earlier findings~\cite{PhysRevA.75.062333, PhysRevA.101.023802}. Figs.~\ref{fig:3}(c) and (d) display the time evolution of the return probability $\mathrm{R_0}(t)$ and the participation function $\mathrm{PR}(t)$ for the same configurations shown in panels (a) and (b), now followed over a longer temporal window. In the travelling self-trapped regime [panel (a) as reference], $\mathrm{R_0}(t)$ decreases continuously over time after an initial transient while $\mathrm{PR}(t)$ saturates at a relatively small value, indicating coherent propagation with limited spatial spreading. In contrast, the soliton-like formations [panel (b) as reference] are characterized by a return probability fluctuating around a finite value and by a small, nearly constant participation function, consistent with confinement around the initial site. Finally, in Figs.~\ref{fig:3}(e) and (f) we analyze $\mathrm{R_0}(t)$ and $\mathrm{PR}(t)$ for the same nonlinearities $\chi_1$ and $\chi_2$ in the presence of spatial [$\theta(n)$] and temporal [$\theta(t)$] inhomogeneities. We observe spatial disorder stabilizing localized patterns, preserving a residual trapped core even in this highly sensitive regime, accompanied by a saturation of the participation function at long times. This behavior contrasts with the weaker-nonlinearity Hadamard-gate scenarios discussed in Fig.~\ref{fig:2}, where the return probability progressively diminishes and the participation function continues to grow under spatial inhomogeneities. Here, stronger nonlinearities counterbalance disorder-induced decoherence, allowing a persistent localized core to survive. Temporal inhomogeneities, on the other hand, act as time-dependent perturbations that continuously disrupt phase coherence, thereby enhancing the emission of dispersive wave-packet components and promoting delocalization and spreading.

To complement the dynamical analysis and provide a global characterization of the asymptotic regimes, we now consider long-time averaged quantities, computed after approximately $7\times 10^4$ time steps, here denoted by $\mathrm{\overline{R_0}}(t_\infty)$ and $\mathrm{\overline{PR}}(t_\infty)$. In Fig.~\ref{fig:4}, these observables are mapped in the $(\chi,\theta_0)$ plane, allowing us to identify the dominant long-time behaviors and to quantify how spatial and temporal inhomogeneities reshape the parameter regions associated with self-trapping. Throughout the figure, black areas correspond to values below the thresholds $\overline{R_0}(t_\infty)=0.03$ and $\overline{PR}(t_\infty)=2$, which are used as operational criteria to delineate regimes where the return probability to the initial site is negligible and the wave packet remains strongly spatially confined, respectively. Moderate variations of these thresholds do not alter the qualitative structure of the resulting phase diagrams. Fig.~\ref{fig:4}(a) and (b) show the homogeneous scenario, where the stationary self-trapped regime becomes noticeable. This regime, characterized by $\mathrm{\overline{R_0}}(t_\infty)\simeq 1$ and $\mathrm{\overline{PR}}(t_\infty)\simeq 1$, becomes predominant in the vicinity of $\theta_0=\pi/2$, in full agreement with Ref.~\cite{PhysRevA.101.023802}. For sufficiently small $\theta_0$, trapped states are absent, and the dynamics remain extended, while travelling soliton-like regimes emerge as $\theta_0$ is tuned around $\pi/4$. In this intermediate region, small values of $\chi$ are associated with orderly propagation of such formations through the lattice, whereas larger nonlinearities drive the system into a regime that is highly sensitive to small variations of $\chi$, signaling the onset of chaotic-like dynamics~\cite{PhysRevA.101.023802}.

Fig~\ref{fig:4}(c) and (d) present the long-time averaged return probability and participation function in the presence of spatial inhomogeneities in the quantum gates, $\theta(n)$. Compared to the homogeneous case, the parameter region associated with stationary self-trapping is substantially reduced. Spatial disorder undermines the phase coherence required for such trapping. As a consequence, $\mathrm{\overline{R_0}}(t_\infty)$ decreases and $\mathrm{\overline{PR}}(t_\infty)$ increases over most of the parameter space, revealing regimes of partial delocalization. In this scenario, stationary self-trapping survives only for values of $\theta_0$ approaching $\pi/2$. The partial delocalization becomes particularly evident near the Hadamard gate configuration ($\theta_0=\pi/4$), underscoring the competition between nonlinearity and spatial disorder. For strong enough nonlinearities, this interplay gives rise to partially localized states, in which a residual trapped core coexists with spatially extended wave-packet components. This phenomenology is consistent with the dynamical behavior reported in Fig.~\ref{fig:2}(c,d) for spatially inhomogeneous systems with $\theta_0=\pi/3$ and $\chi=0.6$, where a finite saturation of the return probability coexists with a growing participation function, indicating the simultaneous presence of a confined core and dispersive spreading. As the nonlinear strength is reduced, this residual core progressively weakens over time and the dynamics become predominantly extended. This trend naturally extends to small gate angles $\theta_0$, where self-trapping is already absent in the homogeneous lattice. In this regime, the long-time averaged return probability remains low while the participation function attains large values, indicating that spatial inhomogeneities do not qualitatively alter the intrinsically delocalized dynamics of the homogeneous framework.

Fig.~\ref{fig:4}(e) and (f) show the corresponding long-time averaged quantities for lattices with temporal inhomogeneities, $\theta(t)$. In this scenario, stationary self-trapping is completely suppressed across the entire $(\chi,\theta_0)$ parameter space. The return probability remains predominantly below the adopted threshold, while the participation function attains large values, suggesting a delocalized wave packet. This overall behavior underscores the effectiveness of temporal disorder in disrupting phase correlations and resonant interference mechanisms that are fundamental to nonlinear trapping. As a result, the long-time dynamics are characterized by an effectively diffusive-like spreading, largely independent of the initial regime observed in the homogeneous lattice.

\section{Summary and concluding remarks}\label{sec:conclusions}

In this work, we investigated the effect of spatial and temporal inhomogeneities on the nonlinear dynamics of quantum walks in one-dimensional lattices. Our analysis focused on how these inhomogeneities of the coin parameters affect, destabilize, or qualitatively transform the nonlinear dynamical behaviors established for homogeneous DTQWs. Building upon the homogeneous lattice as a reference, we observed that spatial inhomogeneities weaken the coherent nonlinear dynamics sustaining both travelling and stationary self-trapped regimes. Such a condition substantially reduces the parameter region associated with robust stationary self-trapping and gives rise to regimes of partial delocalization that are absent in homogeneous lattices. In these regimes, a residual localized core may coexist with dispersive wave-packet components, particularly near the Hadamard gate configuration $\theta_0=\pi/4$, reflecting the competition between nonlinear self-trapping and spatial inhomogeneity. In contrast, for small values of $\theta_0$, the dynamics is predominantly extended, and spatial inhomogeneity does not qualitatively alter the already delocalized behavior. These trends are corroborated by long-time averaged analysis, which reveal a decrease in the return probability accompanied by an increase in the participation function across large regions of parameter space.

A qualitatively distinct behavior emerges when inhomogeneities are introduced in time rather than in space. Temporal inhomogeneities act as time-dependent perturbations that continuously modify the interference conditions of the walk, efficiently suppressing the phase coherence required for nonlinear self-trapping. As a result, both initially travelling and stationary self-trapped states converge toward a common asymptotic regime characterized by effectively diffusive dynamics.

From an experimental perspective, the effects discussed here are directly relevant to a growing number of platforms in which discrete-time quantum walks have been realized, offering independent control over nonlinearity and inhomogeneity. Recent photonic implementations demonstrate the feasibility of nonlinear quantum walks based on measurement-induced or Kerr-like nonlinearities \cite{PhysRevLett.127.163901, NaturePhysics2017}, while independent schemes allow the controlled engineering of spatial and temporal inhomogeneities in interferometric platforms, both integrated and time-multiplexed~\cite{PhysRevLett.106.180403, PhysRevResearch.6.033194, Monika2024, doi:10.1126/sciadv.adh0415, Crespi2013}. Within this context, the contrasting roles identified here for spatial and temporal inhomogeneities suggest experimentally accessible routes to tailor localization, spreading, and transport in discrete-time quantum walk platforms, with potential implications for photonic wave guiding, interference-based control, and soliton-like dynamics \cite{NewJPhys2024_ad1e24}. Further progress combining analytical approaches with dedicated experiments is expected to clarify the interplay between nonlinearity, disorder, and coherence, enabling controlled manipulation of quantum walk dynamics.

\section{Acknowledgments}

This work was partially supported by CAPES (Coordena\c{c}\~ao de Aperfei\c{c}oamento de Pessoal do N\'ivel Superior), CNPq (Conselho Nacional de Desenvolvimento Cient\'ifico e Tecnol\'ogico), and FAPEAL (Funda\c{c}\~ao de Apoio \`a Pesquisa do Estado de Alagoas). Additional support was provided by the project "QuIIN FCRH Mestrado TQ" supported by QuIIN - Quantum Industrial Innovation, EMBRAPII CIMATEC Competence Center in Quantum Technologies, with financial resources from the PPI loT/Manufatura 4.0 of the MCTI grant number 053/2023, signed with EMBRAPII.

\bibliographystyle{nature}
\bibliography{ref}

\end{document}